\documentclass[showpacs,preprintnumbers,amsmath,amssymb,prb,twocolumn,floatfix,amssymb]{revtex4}
\usepackage[dvips]{graphics,color}
\usepackage{dcolumn}
\usepackage{amsmath,amssymb,bbold,bm}
\usepackage[toc]{appendix}
\newcommand{\cev}[1]{\reflectbox{\ensuremath{\vec{\reflectbox{\ensuremath{#1}}}}}}

\begin{document}
\title{Possible charge analogues of spin transfer torques in bulk superconductors}
\author{Ion Garate}
\affiliation{D\'epartement de Physique, Universit\'e de Sherbrooke, Sherbrooke, Qu\'ebec, Canada J1K 2R1}
\date{\today}
\begin{abstract}
Spin transfer torques (STT) occur when electric currents travel through inhomogeneously magnetized systems and are important for the motion of magnetic textures such as domain walls.
Since superconductors are easy-plane ferromagnets in particle-hole (charge) space, it is natural to ask whether any charge duals of STT phenomena exist therein.
We find that the superconducting analogue of the adiabatic STT vanishes in a bulk superconductor with a momentum-independent order parameter, while the superconducting counterpart of the nonadiabatic STT does not vanish. 
This nonvanishing superconducting torque is induced by heat (rather than charge) currents and acts on the charge (rather than spin) degree of freedom. 
It can become significant in the vicinity of the superconducting transition temperature, where it generates a net quasiparticle charge and alters the dispersion and linewidth of low-frequency collective modes.
\end{abstract}
\maketitle

\section {Introduction}
Recent advances in spintronics~\cite{stt} have established an equation that captures the low-energy magnetization dynamics of conducting ferromagnets with smooth magnetic textures:
\begin{equation}
\label{eq:lls}
{\dot{\hat\Omega}}={\bf H}_{\rm eff} \times {\hat \Omega} + {\hat{\Omega}}\,  \times\,{\bar\alpha}\, {\dot{\hat\Omega}} -{\bf v}_T{\cdot}{\boldsymbol\nabla} {\hat\Omega}-{\hat\Omega}\times{\bar \beta} {\bf v}_T\cdot{\boldsymbol\nabla}{\hat\Omega},
\end{equation}
where ${\hat\Omega}$ is the direction of magnetization, ${\dot{\hat\Omega}}=\partial{\hat\Omega}/\partial t$ and ${\bf H}_{\rm eff}$ is a sum of external, anisotropy and exchange fields.
The gyromagnetic ratio has been absorbed into ${\bf H}_{\rm eff}$ so that this quantity has energy units.
Likewise, we set $\hbar=k_B=1$ throughout.
The tensor ${\bar\alpha}=\alpha_{i j}$ is the Gilbert damping  and ${\bf v}_T$ is the ``spin velocity'', proportional to the drift velocity of the quasiparticles under an electric field.
When $v_T=0$, Eq.~(\ref{eq:lls}) is known as the Landau-Lifshitz-Gilbert (LLG) equation.
Transport currents lead to $v_T\neq 0$ and influence the state of non-collinear magnetic systems by exerting a spin transfer torque (STT) on the magnetization: ${\bf v}_T\cdot{\boldsymbol\nabla}{\hat\Omega}$ is known as the adiabatic or Slonczewski STT that results when the spins of current-carrying quasiparticles follow the underlying magnetic landscape; ${\hat \Omega} \times {\bar\beta}{\bf v}_T\cdot{\boldsymbol\nabla}{\hat\Omega}$, where ${\bar\beta}=\beta_{i j}$ is a matrix, is known as the nonadiabatic STT. 

Partly because of its promise for magnetoelectronic applications, and partly because the quantitative description of order parameter manipulation by out-of-equilibrium quasiparticles poses great theoretical challenges, the study of STT has developed into a major research subfield of spintronics. 

The objective of this paper is to translate some of the aforementioned developments to the field of nonequilibrium superconductivity.
It has been long-known~\cite{anderson1958} that a superconductor can be characterized as an XY ferromagnet in charge space, in which electron (hole) degrees of freedom play the role of spin-up (spin-down).
Although this analogy has been fruitfully exploited,\cite{analogues} its emphasis has been placed on the equilibrium properties.\cite{exc}
In fact, the field of nonequilibrium superconductivity flourished, peaked, and was deemed understood without reference to magnetism and before the advent of spintronics and spin torques.\cite{gray1981,langenberg1986,kopnin2001} 
In this paper, we propose the existence of a direct analogue of the adiabatic and nonadiabatic STT in superconductors, and extract some of its physical consequences. 

\section{Landau-Lifshitz equations for superconductivity}

We begin from the effective Hamiltonian describing the states of a conventional s-wave superconductor near the Fermi energy,\cite{wong1988}
\begin{align}
\label{eq:model}
{\cal H}&=\sum_{\bf k}{\hat \Psi}^\dagger_{\bf k} (\xi_{\bf k} \tau^z - \Delta \tau^x){\hat \Psi}_{\bf k}+\sum_{\bf q} u_{\rm imp}({\bf q}) {\hat \rho}_{-\bf q}^z\nonumber\\
 &-\frac{g}{4}\sum_{\bf q} ({\hat \rho}_{\bf q}^x {\hat \rho}_{-\bf q}^x + {\hat \rho}_{\bf q}^y {\hat \rho}_{-\bf q}^y)+\frac{1}{2}\sum_{\bf q} V_{\bf q} {\hat \rho}_{\bf q}^z {\hat \rho}_{-\bf{q}}^z,
\end{align}
where $g$ is the short-range attractive interaction, $V_{\bf q}$ is the long-range Coulomb repulsion (e.g. $V_{\bf q}=4\pi e^2/q^2$ and $V_{\bf q}=2\pi e^2/q$ in three and two dimensions,\cite{units} respectively),  
 $u_{\rm imp}$ is a random non-magnetic disorder potential,
$\hat{\Psi}_{\bf k}=(\psi_{{\bf k}\uparrow},\psi^\dagger_{-{\bf k}\downarrow})$ is the Nambu spinor for spin-up electrons and spin-down holes, $\xi_{\bf k}= k^2/(2 m)-\mu$ is the kinetic energy measured from the Fermi energy $\mu$, $\Delta=g\langle\psi_\uparrow\psi_\downarrow\rangle_{\rm \rm eq}$ is the mean-field (BCS) superconducting gap (chosen to be real and spatially uniform), $\langle\dots\rangle_{\rm eq}$ is the equilibrium expectation value, and $\tau^i$ ($i\in\{x,y,z\}$) are Pauli matrices. 
In addition,
\begin{equation}
\label{eq:rho_op}
{\hat \rho}_{\bf q}^i=\sum_{\bf k}\left[{\hat \Psi}^\dagger_{{\bf k}-{\bf q}}\tau^i{\hat \Psi}_{\bf k}-\langle{\hat \Psi}^\dagger_{{\bf k}-{\bf q}}\tau^i{\hat \Psi}_{\bf k}\rangle_{\rm eq}\right]
\end{equation}
are the generalized density operators associated with amplitude and phase fluctuations of the superconducting order parameter (${\hat \rho}^x$ and ${\hat \rho}^y$, respectively), as well as to charge fluctuations (${\hat \rho}^z$). 
Under a weak external perturbation $V^{\rm ext}$, the density operators in Eq.~(\ref{eq:rho_op}) acquire an expectation value 
\begin{equation}
\label{eq:lr1}
\delta\rho^i({\bf q},\omega)=\chi_{i j}({\bf q},\omega) V_j^{\rm ext}({\bf q},\omega),
\end{equation}
where $\omega$ and ${\bf q}$ are the frequency and wave vector of the perturbation, and a sum over repeated indices is implied. 
The many-body density response function $\chi$ can be conveniently evaluated via $\chi^{-1}=(\chi^{QP})^{-1}-U$, where $U={\rm diag} (g/2,g/2,-V_{\bf q})$ and
\begin{equation}
\label{eq:chi_qp}
\chi_{j j' }^{QP} ({\bf q},\omega) = \sum_{n n'} (f_{n'}-f_n)\frac{\langle n'|\tau^j e^{i {\bf q}\cdot {\bf r}} |n\rangle \langle n|\tau^{j'} e^{-i {\bf q}\cdot {\bf r}} |n'\rangle}{\epsilon_n-\epsilon_{n'}-\omega^+}
\end{equation}
is the quasiparticle (one-body) response function to the sum of external and induced ($U \delta\rho$) perturbation. 
Here, $\epsilon_n$ and $|n\rangle$ are the eigenvalues and eigenvectors of the one-body part of Eq.~(\ref{eq:model}), and $f_n$ is the quasiparticle occupation factor.
Also, $\omega^+=\omega+i 0^+$.
In the limit $V^{\rm ext}\to 0$, the dynamics of order parameter fluctuations follows from 
\begin{equation}
\label{eq:resp}
\left(\begin{array}{ccc} \chi_{xx}^{QP}-\frac{2}{g} & 0 & 0 \\
                            0           &     \chi_{yy}^{QP}-\frac{2}{g} & \chi_{yz}^{QP} \\
                            0           &     \chi_{zy}^{QP}     &  \chi_{zz}^{QP}+\frac{1}{V_{\bf q}}
\end{array}\right)\left(\begin{array}{c} \delta\Delta^x \\ \delta\Delta^y \\ e\phi\end{array}\right)=0,
\end{equation}
where $\delta\Delta^x=(-g/2)\delta\rho^x$ and $\delta\Delta^y=(-g/2)\delta\rho^y$ are order parameter amplitude and phase fluctuations, and $e\phi=V_{\bf q} \delta\rho^z$ is the electrostatic potential energy.
The dispersion $\omega({\bf q})$ of superconducting collective modes is determined from ${\rm det}(\chi^{QP}-U^{-1})=0$.
In Eq.~(\ref{eq:model}) we have set the equilibrium supercurrent to zero. 
Consequently, amplitude fluctuations are decoupled from phase and charge fluctuations in linear response and are unimportant\cite{higgs} for $\omega\ll \Delta$.

In equilibrium (i.e. when $f_n$ is the Fermi distribution), approximate expressions for $\chi^{QP}$ are known both in clean ($\omega\tau\gg 1$)~\cite{wong1988} and disordered~\cite{kulik1981,ohashi1997} superconductors.

Near $T=0$ and for $(\omega, q v_F)\ll\Delta$,  the coupled phase and charge fluctuations obey
\begin{equation}
\label{eq:T_0}
\left(\begin{array}{cc} \frac{\omega^2}{2\Delta^2}- \frac{1}{2\Delta^2}\frac{n_s}{n}\frac{v_F^2 q^2}{d} & i\frac{\omega}{\Delta}\\
                        -i\frac{\omega}{\Delta} & 2+\frac{1}{N_0V_{\bf q}} \\
\end{array}\right)
\left(\begin{array}{c} \delta\Delta^y \\ e\phi\end{array}\right)=0,
\end{equation}
where $N_0$ is the density of states of the normal state at the Fermi energy, $d$ is the dimensionality of the sample, $n$ is the density of electrons and $n_s$ is the $T=0$ superfluid density given by $n_s\simeq n$ for $\Delta\tau\gg 1$ and $n_s\simeq n \pi \Delta\tau$ for $\Delta\tau\ll 1$ ($\tau^{-1}$ is the disorder scattering rate).
The collective mode is an ordinary plasmon with $\omega_\pm ({\bf q})=\pm [2 N_0 v_F^2 V_{\bf q} q^2 (n_s/n d)]^{1/2}$.
In three dimensions, $|\omega_\pm({\bf q})|\gg 2\Delta$ for all ${\bf q}$, thus invalidating Eq.~(\ref{eq:T_0}).
Plasmons with $|\omega_\pm({\bf q})|\ll 2\Delta$ are present in lower dimensions,\cite{mooij1985,buisson1994,camarota2001} where $V_{\bf q}$ diverges more slowly than $q^{-2}$.

It is instructive to rewrite Eq.~(\ref{eq:T_0}) as
\begin{align}
\label{eq:ll_0}
i\omega\delta\rho^y &=-\frac{4\Delta}{g}\left(V_{\bf q}+\frac{1}{2 N_0}\right)\delta\rho^z\nonumber\\
i\omega \delta\rho^z &= \frac{g}{2\Delta} N_0 \frac{n_s}{n} \frac{v_F^2 q^2}{d} \delta\rho^y.
\end{align}
These equations can be viewed as the Landau-Lifshitz equations for a ferromagnet with ``magnetization'' $4\Delta/g$ and an equilibrium orientation along $x$.
The right hand side (r.h.s.) of the first line is the $z$-component of the anisotropy field;~\cite{anis} it originates from the energy cost associated with charge fluctuations and diverges at $q\to 0$ due to the long-range character of Coulomb repulsion. 
The r.h.s. of the second line is the (minus) exchange field, which corresponds to the divergence of the supercurrent.
The $x$- and $y$-components of the anisotropy field vanish, as expected from the $U(1)$ symmetry of the order parameter. 
Damping terms are absent as well because there are no quasiparticles for $T\to 0$ and $\omega\ll 2 \Delta$.
Thus, a superconductor is akin to an insulating, easy-plane ferromagnet.

The superconducting dynamics becomes richer when the number of quasiparticles is significant.
For $T\simeq T_c$ (where $T_c$ is the critical temperature) and $(2\Delta, \tau^{-1})\gg \omega\gg D q^2$, Eq.~(\ref{eq:T_0}) is modified\cite{kulik1981} to
\begin{equation}
\label{eq:T_c}
\left(\begin{array}{cc} \frac{\omega^2 I- v_F^2 q^2 n_s/(n d)}{2\Delta^2}   & i\frac{\omega}{\Delta} I\\
                        -i\frac{\omega}{\Delta} I & 2 I+\frac{2 i D q^2}{\omega}+\frac{1}{N_0 V_{\bf q}} \\
\end{array}\right)
\left(\begin{array}{c} \delta\Delta^y \\ e\phi\end{array}\right)=0,
\end{equation}
where $I=\pi\Delta/(4 T)$ and $D=v_F^2 \tau/d$ is the diffusion constant.
The superfluid density near $T_c$ satisfies $n_s/n\simeq 7 \zeta(3)/(4 \pi^2)\Delta^2/T^2$ for $T_c\tau\gg 1$ and $n_s/n\simeq (\pi/2) (\Delta\tau) \Delta/T$ for $T_c\tau\ll 1$.

In this case, the type of collective mode depends on the magnitude of $\omega/(D q^2)$ relative to $N_0 V_{\bf q}$. 
In 3D, $\omega/(D q^2)\ll N_0 V_{\bf q}$ always 
and Eq.~(\ref{eq:T_c}) yields the Carlson-Goldman (CG) mode:\cite{goldman2006} $\omega_\pm({\bf q}) \simeq \pm (v_{G}^2 q^2-\gamma_G^2)^{1/2} - i\gamma_{G}$, where $v_{G}=v_F\,[n_s/(n I d)]^{1/2}$ and $\gamma_{G}=n_s/(2 n \tau)$ are the velocity and damping of the mode.
In 2D, $\omega/(D q^2)\gg N_0 V_{\bf q}$ can be satisfied at small momenta and therefore a gapless plasmon with $\omega ({\bf q})=\pm (4\pi e^2 n_s/m)^{1/2} q^{1/2}$ emerges in the regime $\omega\ll \gamma_{G}$. 
This mode is replaced by the CG mode when $\omega\gg \gamma_{G}$.

It is again instructive to write Eq.~(\ref{eq:T_c}) in terms of $\delta\rho^i$:
\begin{align}
\label{eq:ll_tc}
i\omega\delta\rho^y &=-\frac{4\Delta}{g}\left(V_{\bf q}+\frac{1}{2 I N_0}+i\frac{D q^2 V_{\bf q}}{I \omega}\right)\delta\rho^z\nonumber\\
i\omega\delta\rho^z &= \frac{g}{2\Delta} N_0 \frac{n_s}{n} \frac{v_F^2 q^2}{d}\delta\rho^y + 2 N_0 D q^2 V_{\bf q}\delta\rho^z.
\end{align}
The first line of Eq.~(\ref{eq:ll_tc}) is essentially the Jospehson relation containing a damping term, which does not have the Gilbert form.
This is because {\em inelastic} scattering processes have been ignored in the derivation of Eq.~(\ref{eq:ll_tc}).
If one incorporates inelastic scattering in the damping term via\cite{kulik1981} $\omega\to \omega+i \tau_E^{-1}$, where $\tau_E^{-1}$ is the inelastic scattering rate, then in the limit $\omega\ll \tau_E^{-1}$  the damping term becomes Gilbert-like with a coefficient
\begin{equation}
\label{eq:alpha}
\alpha_{zz}=\frac{16}{\pi}\frac{V_{\bf q}}{g} (T \tau_E) D q^2 \tau_E.
\end{equation}
Remarkably, $\alpha_{zz}$ is independent of momentum in 3D but it vanishes for $q\to 0$ in lower dimensions.
There are additional peculiarities of Eq.~(\ref{eq:alpha}) compared to what is customary in ferromagnetic metals.
On one hand, although inelastic scattering is acknowledged to be ultimately necessary for magnetization relaxation in conducting ferromagnets, a response function calculation with purely elastic disorder suffices to produce a Gilbert damping term therein.\cite{garate2009}
This is not the case in a superconductor, as evidenced by Eq.~(\ref{eq:ll_tc}). 
On the other hand, Eq.~(\ref{eq:alpha}) is proportional to  $\tau_E^2$, which is neither the conductivity-like nor resistivity-like scaling that one is accustomed to in conducting ferromagnets. 
These differences might be partly reconciled by building a microscopic theory of magnetization damping for insulating ferromagnets near the Curie temperature. 

The second line of Eq.~(\ref{eq:ll_tc}) is the current continuity equation; its last term on the right hand side is the divergence of the quasiparticle current $\sigma{\boldsymbol\nabla}\cdot{\bf E}$, where $\sigma=2 N_0  e^2 D$ is the conductivity and ${\bf E}=-{\boldsymbol\nabla}\phi$ is the electric field. 
In magnetic language, $\sigma{\boldsymbol\nabla}\cdot{\bf E}$ is a Bloch-like relaxation term. 
The reason for $\alpha_{yy}=0$ in the continuity equation can be explained from the breathing Fermi surface picture of magnetism:\cite{kambersky1970}
the energy spectrum is invariant under spatially uniform changes of the phase of the order parameter.
In contrast, changing $\delta\rho^z$ (or $\phi$) modifies the energy spectrum and produces instantaneously-out-of-equilibrium quasiparticle populations, which upon relaxation culminate in $\alpha_{zz}\neq 0$.

\section{Superconducting analogues of spin torques}

So far we have reinterpreted the known dynamics of the superconducting order parameter from the point of view of magnetism.
The response functions discussed above involved quasiparticles in equilibrium with the condensate.
In magnets, transport currents drift quasiparticle populations away from the Fermi distribution, and the ensuing change in the spin response function constitutes the microscopic mechanism for STT.\cite{rossier2004, kohno2006, duine2007}
Next, we search for a dual phenomenon in superconductors.

Departures of the quasiparticle distribution function from equilibrium, $\delta f_{\bf k}$, can be classified according to their parities~\cite{aronov} under ${\bf k}\to -{\bf k}$ and under $\xi_{\bf k} \to -\xi_{\bf k}$.
Here we concentrate on ``transport perturbations'', for which $\delta f_{\bf k} = -\delta f_{-{\bf k}}$.
Neglecting $O(T/\mu)$ terms, transport perturbations that are even (odd) in $\xi_{\bf k}$ induce electrical (heat) currents.
The change in the quasiparticle response function under such perturbation, $\delta\chi^{QP}$, is an odd power of ${\bf q}$ in centrosymmetric superconductors.

We evaluate $\delta\chi^{QP}$ by replacing $|n\rangle $ and $\epsilon_n$ in Eq.~(\ref{eq:chi_qp}) with the eigenvectors and eigenvalues of the clean BCS Hamiltonian in Nambu representation, and by shifting $f_n$ away from the Fermi distribution.
This approximate approach to the full nonlinear response is believed~\cite{duine2007,garate2009} to provide a semi-quantitative microscopic understanding of STT in magnets whose mean free paths are larger than the order parameter coherence length.
Arguably, it only captures the effect of perturbing the quasiparticle distribution function and overlooks the effect of perturbing the quasiparticle  {\em eigenfunctions}.
However, the latter has a parametrically different dependence on $\tau$ and should be subdominant in superconductors with\cite{caveat0} $T_c\tau\gg 1$.
Although $T_c\tau\gg 1$ is a rather restrictive condition, it is still relevant to the dynamics of low-energy collective modes.

\begin{widetext}
A straightforward but delicate computation (see Appendices A and B) gives 
\begin{equation}
\label{eq:yz_1}
\delta\chi_{j j'}^{QP}({\bf q},0)\simeq \delta_{j y}\delta_{j' y} \frac{2\pi i}{q v_F} \sum_{\bf k}\delta f_{\bf k}\frac{|\xi_{\bf k}|}{E_{\bf k}}\left[\delta\left({\hat k}\cdot{\hat q}-\frac{2\xi_{\bf k}}{q v_F}\right)-\delta\left({\hat k}\cdot{\hat q}+\frac{2\xi_{\bf k}}{q v_F}\right)\right]
\end{equation}
\end{widetext}
as the leading nonequilibrium correction to the quasiparticle response in the long-wavelength and low-frequency limit, with $E_{\bf k}=(\xi_{\bf k}^2+\Delta^2)^{1/2}$.
In Eq.~(\ref{eq:yz_1}), the factor multiplying $\delta f_{\bf k}$ is odd under $\xi_{\bf k}\to -\xi_{\bf k}$.
Consequently, to leading order in $T/\mu$, only transport perturbations that are odd under $\xi_{\bf k}\to -\xi_{\bf k}$ can induce $\delta\chi_{j j'}^{QP}\neq 0$.
In other words, perturbations that generate electrical currents do not produce an analogue of STT in particle-hole symmetric superconductors, whereas perturbations that generate thermal currents do.
In direct duality, STT in particle-hole symmetric magnets is induced by electric fields and not by temperature gradients.
Particle-hole asymmetries enable thermally induced STT in magnets and form the basis for spin caloritronics.\cite{bauer2012}
Likewise, in a superconductor, particle-hole asymmetry enables electrically induced analogues of STT; nevertheless, this effect will be relatively very small.

For a uniform temperature gradient, the relaxation time approximation~\cite{aronov} yields
\begin{equation}
\label{eq:rel}
\delta f_{\bf k}\simeq\tau_{\bf k}^s\frac{E_{\bf k}}{T}\frac{\partial f_{\bf k}}{\partial E_{\bf k}}{\bf v}_{\bf k}\cdot{\boldsymbol\nabla}T, 
\end{equation}
where $\tau_{\bf k}^s=\tau E_{\bf k}/|\xi_{\bf k}|$ and ${\bf v}_{\bf k}=\partial E_{\bf k}/\partial {\bf k}= {\bf v}_F \xi_{\bf k}/E_{\bf k}$. 
Upon substituting Eq.~(\ref{eq:rel}) in Eq.~(\ref{eq:yz_1}), we have 
\begin{equation}
\label{eq:yz_3}
\delta\chi_{yy}^{QP}({\bf q},0)\simeq -i N_0 ({\bf q}\cdot{\bf v}_T)/T,
\end{equation}
where 
\begin{equation}
\label{eq:vT}
{\bf v}_T=\frac{\pi D}{4 \cosh^2(\Delta/ 2T)} \frac{{\boldsymbol \nabla}T}{T}
\end{equation}
is the superconducting dual to the ``spin velocity''.
Illustrating the fact that the superconducting STT emerges from the interplay between the order parameter and quasiparticles, $v_T\propto \exp(-\Delta/T)$ when $T\ll \Delta$.
At the same time, it will be apparent below that the influence of $v_T$ on the superconducting dynamics vanishes when $T\to T_c$.
Hence, $T\simeq\Delta$ is the optimal temperature to maximize the superconducting STT.
For $T\gtrsim\Delta$, one has $v_T/v_F \sim (l/L) \delta T/T$, where $L$ is the linear dimension of the sample, $l$ is the elastic mean free path and $\delta T$ is the temperature difference between the ends of the sample. 
Taking $\delta T/T\simeq 0.01$ and $L\simeq 10^3 l$, it follows that $v_T\simeq 10^{-5} v_F$.

Equation~(\ref{eq:yz_1}) is unusual from the point of view of magnetism.
On one hand,  $\delta\chi_{y z}^{QP}({\bf q},0)=0$ implies that there is no superconducting counterpart of the adiabatic STT.
% (at least within our approximate approach).
As shown in Appendix A, this result emerges from a perfect cancellation between interband and intraband contributions ($n\neq n'$ and $n=n'$ terms in Eq.~(\ref{eq:chi_qp}), respectively), each of which are nonzero in presence of a temperature gradient.
Such cancellation, which has not been found to occur in ordinary ferromagnets, holds regardless of the temperature and crucially relies on the momentum-independence of the simple BCS gap.
For a momentum-dependent gap, we instead find
\begin{equation}
\label{eq:delta}
\delta\chi_{y z}^{QP}({\bf q},0)=2 i\sum_{\bf k} \delta f_{\bf k}\frac{\Delta_{\bf k}-\Delta_{{\bf k-q}}}{E_{\bf k}^2-E_{\bf k-q}^2} -({\bf q}\to -{\bf q}),
\end{equation}
which implies that a superconducting analogue of the adiabatic STT can occur in unconventional superconductors.
The evaluation of Eq.~(\ref{eq:delta}) for different types of order parameters and transport perturbations is a potentially interesting problem that will be addressed elsewhere.
For the remainder of this paper, we restrict ourselves to a momentum-independent gap. 

Another peculiarity of Eq.~(\ref{eq:yz_1}) is ${\rm Im}\,\delta\chi^{QP}_{yy}({\bf q},0)\propto {\bf q}\cdot\boldsymbol\nabla T$ and ${\rm Im}\,\delta\chi_{zz}^{QP}({\bf q},0)= 0$, which means that a superconducting analogue of the nonadiabatic STT exists with $\beta_{yy}\neq 0$ and $\beta_{zz}=0$.
The presence of a nonadiabatic STT in absence of an adiabatic STT is unheard of in ordinary ferromagnets.
Finally, ${\rm Re}\,\delta\chi^{QP}_{\perp\perp}({\bf q},0)=0$ is a consequence of inversion symmetry and has a well-understood correspondence in magnetism: transport currents do not modify the anisotropy field of centrosymmetric magnets.
For a centrosymmetric superconductor, the leading reactive (real) terms in $\delta\chi^{QP}$ appear when $\omega\neq 0$ and are evaluated in Appendix C. 
These contributions arise because the kinetic energy term in Eq.~(\ref{eq:model}) acts like a momentum-dependent magnetic field. 
Analogous terms in centrosymmetric magnets with spin-orbit interactions are commonly neglected in the low-frequency and long-wavelength expansion.

After taking Eq.~(\ref{eq:yz_3}) into account, and having verified (cf. Appendix D) that amplitude fluctuations remain decoupled from phase/charge fluctuations in presence of a temperature gradient,  Eq.~(\ref{eq:T_c}) is generalized to
\begin{equation}
\label{eq:T_c2}
\left(\begin{array}{cc} \frac{\omega^2 I-  \frac{v_F^2 q^2 n_s}{n d}}{2\Delta^2}+\delta\chi_{yy}^{QP}   & i\frac{\omega}{\Delta} I\\
                        -i\frac{\omega}{\Delta} I & 2 I+\frac{2 i D q^2}{\omega}+\frac{1}{N_0 V_{\bf q}} \\
\end{array}\right)
\left(\begin{array}{c} \delta\Delta^y \\ e\phi\end{array}\right)=0.
\end{equation}
The corresponding generalization of Eq.~(\ref{eq:ll_tc}) is
\begin{align}
\label{eq:rhola}
&i\omega\delta\rho^y =-\frac{4\Delta}{g}\left(V_{\bf q}+\frac{1}{2 I N_0}+i\frac{D q^2 V_{\bf q}}{I \omega}\right)\delta\rho^z\nonumber\\
&i\omega\delta\rho^z = \frac{g}{2\Delta} N_0 \frac{n_s}{n} \frac{v_F^2 q^2}{d}\delta\rho^y + \sigma{\boldsymbol \nabla}\cdot{\bf E}+ \frac{\partial \delta\rho^z}{\partial t}\Big|_{\rm STT}.
\end{align}
The last term of Eq.~(\ref{eq:rhola}),
\begin{equation}
\label{eq:torque}
\frac{\partial\delta\rho^z}{\partial t}\Big|_{\rm STT}=i g N_0 \frac{\Delta}{T} ({\bf q}\cdot{\bf v}_T)\delta\rho^y,
\end{equation}
is a nonadibatic torque induced by a combination of a supercurrent and a temperature gradient [the superfluid momentum is ${\bf P}=-({\boldsymbol\nabla}\delta\Delta^y)/(2\Delta)\to i g \,{\bf q}\,\delta\rho^y/(4 \Delta)$].
The idea that a temperature gradient and a spatially uniform supercurrent can conspire to generate a net quasiparticle charge (also known as ``quasiparticle charge imbalance'') is not new.\cite{pethick1979,schmid1979,clarke1979}
Here we have derived a dynamical version of a similar result from an alternative viewpoint, without assuming a uniform equilibrium supercurrent, and have identified it as a manifestation of the superconducting STT.

Next, we evaluate the influence of the superconducting STT on low-energy collective modes, which appears to have remained unexplored in the literature.
The magnetic counterpart of this effect is known to be important.\cite{rossier2004}
Since we have calculated the STT term for $\omega\to 0$, it is legitimate to question whether Eq.~(\ref{eq:yz_1}) is applicable to collective modes.
The answer is affirmative provided that $\omega\ll q v_F$, because $\delta\chi^{QP}({\bf q},\omega)\simeq \delta\chi^{QP}({\bf q},0)$ under this condition.\cite{contrast}

With this proviso, let us begin from the CG mode, for which both $\omega/(D q^2)\ll N_0 V_{\bf q}$ and $\omega\ll q v_F$ are readily satisfied.
When $q\gg \gamma_{G}/v_{G}\equiv q_G$, Eq.~(\ref{eq:T_c2}) yields  $\omega_\pm({\bf q})\simeq \pm v_G q-i\gamma_G+\delta\omega_\pm({\bf q})$, where 
\begin{equation}
\label{eq:cgt}
\delta\omega_\pm ({\bf q}) \simeq \pm \left(1+i\frac{\gamma_G}{q v_G}\right)\frac{\pi}{4}\frac{\Delta^2}{T^2} \frac{{\bf q}\cdot {\boldsymbol\nabla }T}{q^2}.
\end{equation}
For $|{\boldsymbol \nabla} T|/T=10^3$ m$^{-1}$ (which should be achievable in mesoscopic samples) and $v_F=5\times 10^5 {\rm m/s}$, we obtain $|{\rm Re}\,\delta\omega_\pm|\simeq (\Delta/T_c)^{1/2} T_c \tau\, (q_G/q)\,{\hat q}\cdot{\hat n} [{\rm GHz}]$, where ${\hat n}$ is the direction of the temperature gradient.
For $T_c\tau\gg 1$ (which is the regime for which we have calculated the superconducting STT), this shift can exceed $\gamma_G$, and thus be observable.   
When $v_T=0$, the ordinary CG mode becomes overdamped at $q<q_G$.
However, $v_T\neq 0$ introduces a characteristic momentum, $q^*=\tau {\hat q}\cdot{\boldsymbol\nabla}T$, 
below which a propagating mode reappears\cite{other} with an anomalous dispersion
\begin{equation}
\label{eq:cube}
\omega({\bf q}) \simeq \frac{7 \zeta(3)}{4 \pi^3}\frac{v_G^2 q^2}{\gamma_G}\frac{q}{q^*},\,\mbox{   ($q\ll |q^*|$)}.
\end{equation}
Note that $q^*\in[-\tau |{\boldsymbol\nabla}T|, +\tau |{\boldsymbol\nabla}T|]$ as a function of the angle between ${\bf q}$ and ${\boldsymbol\nabla}T$. 
When $|{\boldsymbol \nabla} T|/T=10^3 {\rm m}^{-1}$, $q^*\simeq 10^3 \,T_c\tau\, {\hat q}\cdot{\hat n}[{\rm m}^{-1}]$ can be of the order of $q_G$.
Because of its $q^3$ scaling, Eq.~(\ref{eq:cube}) is compatible with $\omega\gg D q^2$ only if $T_c/\Delta\gtrsim 10$, i.e. exceedingly close to $T_c$. 

The influence of the superconducting STT can also be significant on the gapless plasmon modes that exist for $\omega\ll\gamma_G$ in lower dimensional systems.
For example, in a 2D superconductor, the modified plasmon dispersion at $T\lesssim T_c$ reads
\begin{equation}
\label{eq:plt}
\omega_\pm ({\bf q})\simeq \pm \sqrt{\frac{4 \pi e^2 n_s}{m} q + i 8\pi I N_0 e^2 D \Delta\frac{{\hat q}\cdot{\boldsymbol\nabla}T}{T}},
\end{equation}
where we have omitted a subleading term that originates from Appendix C and changes the real part of the dispersion. 
In this case, the requirement $\omega\ll (q v_F,\gamma_G)$ is rather restrictive: Eq.~(\ref{eq:plt}) is applicable if $e^2 N_0 n_s/n\ll q\ll n_s/(e^2 N_0 l^2 n)$.
This condition is compatible with $T_c\tau>1$ only if $T_c[{\rm meV}] \epsilon[10^4]/v_F[10^5 {\rm m/s}]> 0.5$, where $\epsilon$ is the dielectric constant in units of the vacuum permittivity and we have assumed that the effective electron mass agrees with its value in vacuum.
For a large dielectric constant\cite{buisson1994} of $\epsilon\simeq 2\times 10^4$ and the aforementioned values of parameters, 
the bare plasmon frequency of $\simeq 0.3 q^{1/2} [{\rm m}^{-1/2}](\Delta/T_c)[{\rm GHz}]$ is accompanied by a STT-induced linewidth of $\simeq 0.5 |q^*[{\rm m}^{-1}]|^{1/2} (\Delta/T_c) [{\rm GHz}]$ in Eq. ~(\ref{eq:plt}).
It follows that the 2D plasmon gets overdamped at $q \lesssim |q^*|$.

In Ref.~[\onlinecite{buisson1994}], the authors were able to measure the superconducting plasmon frequency with an accuracy of $\pm 1{\rm MHz}$. 
With such a resolution,\cite{caveat} the STT-induced linewidth should be observable in mesoscopic samples (where the total temperature drop across the sample under $|{\boldsymbol\nabla}T|/T\simeq 10^3 {\rm m}^{-1}$ is a small fraction of the sample temperature).
%However, it is plausible that superconducting torque effects are observable in the dirty limit as well. 
In sum, perhaps unexpectedly,\cite{ohashi1998} the plasmon dispersion is affected in the superconducting phase when the quasiparticles are driven out of equilibrium by a temperature gradient. 

When $\omega\gg v_F q$, the superconducting analogue of the nonadiabatic STT vanishes (much like the usual Landau damping vanishes in the same regime) and $\delta\chi^{QP}_{y z}({\bf q},\omega)$ is no longer zero. 
Therefore, in this case, the leading influence of a transport perturbation in the 2D plasmon dispersion originates from reactive terms: the outcome is similar to the one described in Appendix E for a clean superconductor (modulo replacing $n$ by $n_s$).

\section{Discussion}

The two lines of Eq.~(\ref{eq:T_c2}) coincide with the time-dependent Ginzburg-Landau (TDGL) equations derived from the kinetic theory approach,\cite{ambegaokar1979} so long as one takes $v_T=0$ and $1/V_{\bf q}=0$ in the former and $\omega\tau_E\gg 1$ in the latter. 
Neglecting  $1/V_{\bf q}$ in Ref.~[\onlinecite{ambegaokar1979}] was appropriate for the study of low-energy dynamics of 3D superconductors near $T_c$; however, it must be retained in order to capture the gapless plasmon modes of lower dimensional systems.

Often, the regime of interest for applications of the TDGL equations is $\omega\ll\tau_E^{-1}$. 
Since we have neglected inelastic scattering processes in $\chi^{QP}$ (except for a brief interlude in the discussion of damping), we cannot make any rigorous statements in this regime. 
However, we extrapolate Eq.~(\ref{eq:T_c2}) to $\omega \tau_E\ll 1$ according to the prescription of Ref.~[\onlinecite{kulik1981}] and immediately arrive at
\begin{align}
\label{eq:tdgl}
-\frac{I}{\tau_E}\left(e\phi-i\omega\frac{\delta\Delta^y}{2\Delta}\right) &=\frac{n_s}{n}\frac{v_F^2 q^2}{d}\frac{\delta\Delta^y}{2\Delta} + i \frac{\Delta}{T} ({\bf q}\cdot{\bf v}_T) \delta\Delta^y\nonumber\\
-I\left(e\phi-i\omega\frac{\delta\Delta^y}{2\Delta}\right) &=D q^2 \tau_E e\phi,
\end{align}
having neglected $1/V_{\bf q}$ in the second line. 
The combination of the electrostatic potential and the time derivative of the superconducting phase appearing on the left hand side of Eq.~(\ref{eq:tdgl}) is variously referred to as the gauge-invariant potential or the condensate chemical potential. 
Near $T_c$ it approximately coincides with the difference between the quasiparticle and condensate electrochemical potentials, which in turn is proportional to the quasiparticle charge imbalance.\cite{kadin}
%In addition, the quasiparticle charge imbalance is proportional to this difference between the electrochemical potentials.
The second line in Eq.~(\ref{eq:tdgl}) yields the steady-state penetration depth of an electric field into a superconductor; it remains unchanged in presence of a transport perturbation.
In sum, Eq.~(\ref{eq:tdgl}) agrees with the appropriate version of Ref.~[\onlinecite{ambegaokar1979}], insofar as $v_T=0$.
Thus, the superconducting STT term in Eq.~(\ref{eq:tdgl}) appears to modify the existing TDGL theory somewhat like the STT terms in Eq.~(\ref{eq:lls}) modify the LLG equations.

Nonetheless, it must be mentioned that a term similar to the STT in the first line of Eq.~(\ref{eq:tdgl}) has been derived using the kinetic theory approach, both in the clean and dirty limits.\cite{schon1986}
This term was discussed only for the steady state and for a spatially uniform supercurrent; no observations were made about its influence in the dynamics (e.g. collective modes).
A possible reason for this is that the effect of the superconducting STT in the collective modes is small for macroscopic superconductors.
More so, at the time of Ref.~[\onlinecite{schon1986}] it was unfeasible to contemplate connections between superconductivity and spin torques.

In Ref.~[\onlinecite{schon1981}], an additional term proportional to ${\bf P}\cdot {\bf E}$ was proposed phenomenologically for the first line of Eq.~(\ref{eq:tdgl}). 
As shown in Appendix B, our theory indicates that the coefficient multiplying such term is nonzero only due to particle-hole asymmetry.
Finally, to the best of our knowledge, the conventional equations of motion for nonequilibrium superconductivity do not include an analogue of the adiabatic STT, which according to Eq.~(\ref{eq:delta}) can exist in superconductors with a momentum-dependent gap.

Why does the adiabatic torque vanish (via a nontrivial cancellation) in a superconductor with a momentum-independent order parameter?
In presence of an adiabatic STT, the instantaneous quasiparticle charge imbalance would follow ${\bf P}\cdot{\boldsymbol\nabla T}$ adiabatically. 
However, this would be unphysical unless there was a relaxation mechanism for the charge imbalance. 
It turns out that in absence of inelastic scatterers, magnetic impurities and equilibrium supercurrents, a momentum-dependent gap (in conjunction with elastic disorder) is the only way to relax the quasiparticle charge imbalance.\cite{clark}
This, we speculate, may be behind the cancellation of the adiabatic STT in our approach.

\section{Conclusions}

Motivated by recent advances in the understanding of spin torques in magnetic systems, we have revived a known mathematical correspondence between ferromagnetism and superconductivity in order to reinterpret the dynamics of a superconducting order parameter from a ``spintronics point of view''.
This approach has enabled us to suggest a nonequilibrium superconducting effect that is dual to the nonadiabatic spin transfer torque (STT) of magnetic systems. 
This ``torque'' acts on the charge degree of freedom, is induced mainly by temperature gradients, and has its largest magnitude in the vicinity of the transition temperature.
In contrast, the adiabatic STT of ferromagnets appears to have a superconducting counterpart only if the order parameter is momentum-dependent (cf. Eq.~(\ref{eq:delta})).
These results have been derived from linear response theory with respect to the transport steady state.
Although less accurate and general than the full nonlinear response theory, our approach is considerably simpler and is expected to provide the correct qualitative picture in clean superconductors at frequencies that exceed the inelastic scattering rates.

The superconducting torque we have identified is behind a known thermoelectric effect, and leads to hitherto unpredicted changes in the dispersion of collective modes.
It remains to be seen whether the superconducting torque will be effective in altering the configuration of inhomogeneous order parameter textures (such as vortices and phase-slip centers) at the meso- and nanoscale.
It will also be useful to explore the spin torque analogues in Josephson junction arrays, as well as in unconventional superconductors with and without inversion symmetry.

\acknowledgements
I am indebted to A.H. MacDonald for inspiring me to think about this problem.
In addition, I thank A.H. MacDonald,  T. Pereg-Barnea, B. Reulet, D. S\'en\'echal and A.-M. Tremblay for interesting questions and comments.
This project started at the University of Texas at Austin and has subsequently received financial support from Yale University, Universit\'e de Sherbrooke, and Canada's NSERC.

\begin{widetext}

%\begin{appendices}
\appendix

\section {Superconducting analogue of the adiabatic STT}

As indicated in the main text, in order to evaluate the change of the quasiparticle response functions under a transport perturbation, we compute Eq.~(\ref{eq:chi_qp}) using the eigenstates and eigenvalues of a clean superconductor, and shift the quasiparticle distributions away from equilibrium.
The eigenvalues are $E_{{\bf k}+}=(\xi_{\bf k}^2+\Delta^2)^{1/2}\equiv E_{\bf k}$ and $E_{{\bf k} -}=-E_{\bf k}$, and the corresponding eigenvectors read
\begin{equation}
|{\bf k} +\rangle = \left(\begin{array}{c} -\sin\frac{\theta_{\bf k}}{2} \\ \cos\frac{\theta_{\bf k}}{2}\end{array}\right) \mbox{   ;   } |{\bf k} -\rangle = \left(\begin{array}{c} \cos\frac{\theta_{\bf k}}{2} \\ \sin\frac{\theta_{\bf k}}{2}\end{array}\right),
\end{equation}
where $\cos\theta_{\bf k}=\xi_{\bf k}/E_{\bf k}$.

The adiabatic STT appears at first order in $q$ (i.e. first spatial derivative, cf. Eq.~(\ref{eq:lls})) and zeroth order in $\omega$.
Hence we concentrate on the small-momentum expansion of
\begin{equation}
\chi_{yz}^{QP}({\bf q},0)=\sum_{{\bf k}\alpha\beta} (f_{{\bf k+q}\alpha}-f_{{\bf k}\beta})\frac{\langle {\bf k+q}\alpha|\tau^y|{\bf k}\beta\rangle\langle{\bf k}\beta|\tau^z|{\bf k+q}\alpha\rangle}{E_{{\bf k}\beta}-E_{{\bf k+q}\alpha} - i 0^+}.
\end{equation}
Noting that $\langle {\bf k}'\alpha|\tau^y|{\bf k}\beta\rangle\langle{\bf k}\beta|\tau^z|{\bf k}'\alpha\rangle$ is purely imaginary, we write 
\begin{equation}
\chi_{y z}^{QP}({\bf q},0)=-\sum_{{\bf k}\alpha\beta} f_{{\bf k}\alpha}\left[\frac{\langle {\bf k}\alpha|\tau^y|{\bf k-q}\beta\rangle\langle{\bf k-q}\beta|\tau^z|{\bf k}\alpha\rangle}{E_{{\bf k}\alpha}-E_{{\bf k-q}\beta}+ i 0^+}-\left(\begin{array}{c} 0^+\to 0^-\\ {\bf q}\to -{\bf q}\end{array}\right)\right].
\end{equation}
Then,
\begin{equation}
{\rm Re}\,\chi_{y z}^{QP}({\bf q},0)=i \pi\sum_{{\bf k}\alpha\beta} f_{{\bf k}\alpha}\left[\langle {\bf k}\alpha|\tau^y|{\bf k-q}\beta\rangle\langle{\bf k-q}\beta|\tau^z|{\bf k}\alpha\rangle\delta(E_{{\bf k}\alpha}-E_{{\bf k-q}\beta}) + ({\bf q}\to -{\bf q})\right]
\end{equation}
The quantity inside the square brackets is even under ${\bf q}\to -{\bf q}$, which implies that it is even under ${\bf k}\to -{\bf k}$ as well.
Accordingly, the real part of $\chi_{yz}^{QP}({\bf q},0)$ remains zero in presence of a transport perturbation.

Hereafter we focus on the imaginary part,
\begin{equation}
{\rm Im}\,\chi_{y z}^{QP}({\bf q},0)=-\sum_{{\bf k}\alpha\beta} f_{{\bf k}\alpha}\left[\frac{{\rm Im}[\langle {\bf k}\alpha|\tau^y|{\bf k-q}\beta\rangle\langle{\bf k-q}\beta|\tau^z|{\bf k}\alpha\rangle]}{E_{{\bf k}\alpha}-E_{{\bf k-q}\beta}}-({\bf q}\to -{\bf q})\right]
\end{equation}
Let us separate the intraband and interband contributions as ${\rm Im} \chi_{yz}^{QP} = {\rm Im} \chi_{yz}^{\rm intra} + {\rm Im} \chi_{yz}^{\rm inter}$.
First, we consider the intraband part:
\begin{equation}
{\rm Im}\chi_{yz}^{\rm intra}({\bf q},0) \simeq -\sum_{{\bf k}\alpha} f_{{\bf k}\alpha}\left[\frac{{\rm Im}[\langle{\bf k}\alpha|\tau^y|{\bf k-q} \alpha\rangle\langle{\bf k-q}\alpha|\tau^z|{\bf k}\alpha\rangle]}{{\bf v}_{{\bf k}\alpha}\cdot{\bf q}-\frac{1}{2}({\bf q}\cdot{\boldsymbol\nabla}_{\bf k})^2 E_{{\bf k}\alpha}} - ({\bf q}\to -{\bf q})\right],
\end{equation}
where ${\bf v}_{{\bf k}\alpha}=\partial E_{{\bf k}\alpha}/\partial {\bf k}$.
Although the second term in the denominator is of higher order in $q$ than the first term, it cannot be neglected because it eventually makes a $\sim O(q)$ contribution to ${\rm Im}\chi_{yz}^{\rm intra}$.
 
Making a long wavelength expansion of the overlap matrix elements, and noting that $\langle{\bf k}\alpha|\tau^y|{\bf k}\alpha\rangle= 0$, we have
\begin{align}
{\rm Im}\chi_{yz}^{\rm intra}({\bf q},0) &\simeq -2 \sum_{{\bf k}\alpha} f_{{\bf k}\alpha} {\rm Im}\left[\frac{\langle{\bf k}\alpha|\tau^y {\bf q}\cdot{\vec\nabla}_{\bf k}|{\bf k}\alpha\rangle \langle{\bf k}\alpha| {\bf q}\cdot{\cev\nabla}_{\bf k} \tau^z|{\bf k}\alpha\rangle}{{\bf v}_{{\bf k}\alpha}\cdot {\bf q}}
+\frac{1}{2}\frac{\langle{\bf k}\alpha|\tau^y ({\bf q}\cdot{\vec\nabla}_{\bf k})^2|{\bf k}\alpha\rangle \langle {\bf k}\alpha|\tau^z|{\bf k}\alpha\rangle}{{\bf v}_{{\bf k}\alpha}\cdot{\bf q}}\right.\nonumber\\
&~~~~~~~~~~~~~~~~~~~~~~-\left.\frac{1}{2}\frac{({\bf q}\cdot{\vec\nabla}_{\bf k})^2 E_{{\bf k}\alpha}}{({\bf v}_{{\bf k}\alpha}\cdot{\bf q})^2}\langle {\bf k}\alpha|\tau^y {\bf q}\cdot{\vec\nabla}_{\bf k}|{\bf k}\alpha\rangle\langle{\bf k}\alpha|\tau^z|{\bf k}\alpha\rangle\right],
\end{align}
where ${\vec\nabla}_{\bf k} = \partial/\partial {\bf k}$ acting on the right and ${\cev\nabla}_{\bf k} = \partial/\partial {\bf k}$ acting on the left.
Computing the matrix elements, we get
\begin{equation}
{\rm Im}\chi_{y z}^{\rm intra}({\bf q},0)=-\frac{1}{2}\sum_{\bf k} (f_{{\bf k}+}+f_{{\bf k}-})\frac{\cos\theta_{\bf k} ({\bf q}\cdot{\vec\nabla}_{\bf k})^2\theta_{\bf k}}{{\bf v}_{\bf k}\cdot {\bf q}}
\end{equation}
where we have used ${\bf v}_{{\bf k}+}=-{\bf v}_{{\bf k}-}=\partial E_{\bf k}/\partial{\bf k}\equiv {\bf v}_{\bf k}$.
In addition, we have relied on $\xi_{\bf k-q}\simeq \xi_{\bf k}-{\bf q}\cdot{\bf v}_F$, and have verified that the omission of the $q^2/(2 m)$ term does not change the final results.
Evidently ${\rm Im}\chi_{yz}^{\rm intra}({\bf q},0)=0$ in equilibrium.
%Next, we note that $\partial_{k_i}\theta_{\bf k}=-\Delta v_{F,i}/E_k^2$ is even under $\xi_{\bf k} \to -\xi_{\bf k}$, whereas $\partial_{k_i}\partial_{k_j}= 2 \Delta \xi_{\bf k} v_{F i} v_{F j}/E_k^4$ is odd (note that we have approximated $\xi\simeq v_F (k-k_F)$, which is justified for the low energy dynamics of a superconductor at temperatures $\leq T_c$).
%In addition, $\cos\theta=\xi/E_k$ and ${\bf v}_{\bf k}\simeq (\xi_{\bf k}/E_{\bf k}) {\bf v}_F$ are odd under $\xi_{\bf k} \to -\xi_{\bf k}$ and $\sin\theta_{\bf k}=\Delta/E_k$ is even.
%Consequently, it follows that only odd in $\xi$ perturbations will result in a change of ${\rm Im}\chi_{yz}^{\rm intra}$.
%This means that an electric field or a supercurrent will not produce any effect, whereas a temperature gradient will.
For transport perturbations, one has $\delta f_{{\bf k},+} = \delta f_{{\bf k},-}\equiv \delta f_{\bf k}$.
For example, the charge and heat quasiparticle currents are given by
\begin{align}
{\bf j}_e=& \sum_{{\bf k}\alpha} q_{{\bf k}\alpha} {\bf v}_{{\bf k}\alpha} \delta f_{{\bf k}\alpha}=\sum_{\bf k} e (\xi_{\bf k}/E_{\bf k})^2 {\bf v}_F (\delta f_{{\bf k}+}+\delta f_{{\bf k}-})=2\sum_{\bf k} e (\xi_{\bf k}/E_{\bf k})^2 {\bf v}_F \delta f_{\bf k} \nonumber\\
{\bf j}_h=&\sum_{{\bf k}\alpha} {\bf v}_{{\bf k}\alpha} E_{{\bf k}\alpha} \delta f_{{\bf k}\alpha}=\sum_{\bf k} {\bf v}_F \xi_{\bf k} (\delta f_{{\bf k}+}+\delta f_{{\bf k}-})=2 \sum_{\bf k} {\bf v}_F \xi_{\bf k} \delta f_{\bf k}, 
\end{align}
where we have used $q_{{\bf k}\alpha}=\langle {\bf k}\alpha|e \tau^z|{\bf k}\alpha\rangle=e \xi_{\bf k}/E_{{\bf k}\alpha}$ as the quasiparticle charge. 
Incidentally, these expressions reflect the fact that, in presence of particle-hole symmetry,  a transport perturbation that is even (odd) under $\xi_{\bf k}\to -\xi_{\bf k}$ generates an electric (heat) current.

Consequently,
\begin{equation}
{\rm Im}\chi_{y z}^{\rm intra}({\bf q},0)=
%\sum_{\bf k} \delta f_{\bf k}\frac{\sin\theta_{\bf k} ({\bf q}\cdot{\vec\nabla}_{\bf k}\theta_{\bf k})^2-\cos\theta_{\bf k} ({\bf q}\cdot{\vec\nabla}_{\bf k})^2\theta_{\bf k}}{{\bf v}_{\bf k}\cdot {\bf q}}
-2\sum_{\bf k}\delta f_{\bf k}\frac{\xi_{\bf k}\Delta}{E_{\bf k}^4} {\bf v}_F\cdot {\bf q}
\end{equation}
To order $O(T/\mu)$, only transport perturbations that are odd under $\xi_{\bf k}\to -\xi_{\bf k}$ contribute to ${\rm Im}\delta\chi_{y z}^{\rm intra}$.

Next, we compute the interband contribution
\begin{equation}
{\rm Im}\chi_{yz}^{{\rm inter}}({\bf q},0)=-\sum_{{\bf k},\alpha\neq\beta} f_{{\bf k},\alpha}\left[\frac{{\rm Im}[\langle {\bf k}\alpha|\tau^y|{\bf k}-{\bf q}\beta\rangle\langle{\bf k-q}\beta|\tau^z|{\bf k}\alpha\rangle]}{E_{{\bf k}\alpha}-E_{{\bf k}-{\bf q}\beta}} - ({\bf q}\to - {\bf q})\right].
\end{equation}
Expanding the denominator to leading order in {\bf q},
\begin{equation}
{\rm Im}\chi_{yz}^{{\rm inter}}({\bf q},0)\simeq -\sum_{{\bf k},\alpha\neq\beta} f_{{\bf k}\alpha} \frac{{\rm Im}[\langle{\bf k}\alpha|\tau^y|{\bf k}-{\bf q}\beta\rangle\langle{\bf k}-{\bf q}\beta|\tau^z|{\bf k}\alpha\rangle]}{E_{{\bf k}\alpha}-E_{{\bf k}\beta}}\left(1-\frac{{\bf v}_{{\bf k}\beta}\cdot{\bf q}}{E_{{\bf k}\alpha}-E_{{\bf k}\beta}}\right) - ({\bf q}\to -{\bf q}) \equiv A + B,
\end{equation}
where
\begin{equation}
A\equiv -\sum_{{\bf k},\alpha\neq\beta} f_{{\bf k}\alpha}\frac{{\rm Im}[\langle{\bf k}\alpha|\tau^y|{\bf k}-{\bf q}\beta\rangle\langle{\bf k}-{\bf q}\beta|\tau^z|{\bf k}\alpha\rangle]}{E_{{\bf k}\alpha}-E_{{\bf k}\beta}} - ({\bf q}\to -{\bf q})
\end{equation}
and
\begin{equation}
B\equiv \sum_{{\bf k},\alpha\neq\beta} f_{{\bf k}\alpha}\frac{{\rm Im}[\langle{\bf k}\alpha|\tau^y|{\bf k}-{\bf q}\beta\rangle\langle{\bf k}-{\bf q}\beta|\tau^z|{\bf k}\alpha\rangle]}{(E_{{\bf k}\alpha}-E_{{\bf k}\beta})^2} {\bf v}_{{\bf k}\beta}\cdot{\bf q} - ({\bf q}\to -{\bf q}).
\end{equation}
Expanding the matrix elements, 
\begin{equation}
A\simeq  2\sum_{{\bf k},\alpha\neq\beta} f_{{\bf k}\alpha}\left[\frac{{\rm Im}(\langle{\bf k}\alpha|\tau^y {\bf q}\cdot {\vec\nabla}_{\bf k}|{\bf k}\beta\rangle\langle{\bf k}\beta|\tau^z|{\bf k}\alpha\rangle)}{E_{{\bf k}\alpha}-E_{{\bf k}\beta}}+\frac{{\rm Im}(\langle{\bf k}\alpha|\tau^y|{\bf k}\beta\rangle\langle{\bf k}\beta|{\bf q}\cdot{\cev\nabla}_{\bf k}\tau^z|{\bf k}\alpha\rangle)}{E_{{\bf k}\alpha}-E_{{\bf k}\beta}}\right],
\end{equation}
and thus
\begin{equation}
A \simeq -\sum_{\bf k} (f_{{\bf k}+}+ f_{{\bf k},-})\frac{\cos\theta_{\bf k} ({\bf q}\cdot{\vec\nabla}_{\bf k}\theta_{\bf k})}{2 E_{\bf k}}=\sum_{\bf k}\delta f_{\bf k}\frac{\xi_{\bf k}\Delta}{E_{\bf k}^4} {\bf v}_F\cdot{\bf q}.
\end{equation}
Similarly,
\begin{equation}
B\simeq 2 \sum_{{\bf k},\alpha\neq\beta} f_{{\bf k}\alpha}\frac{{\rm Im}(\langle{\bf k}\alpha|\tau^y|{\bf k}\beta\rangle\langle{\bf k}\beta|\tau^z|{\bf k}\alpha\rangle)}{(E_{{\bf k}\alpha}-E_{{\bf k}\beta})^2} {\bf v}_{{\bf k}\beta}\cdot{\bf q} = \sum_{\bf k}\delta f_{{\bf k}} \frac{\xi_{\bf k}\Delta}{E_{\bf k}^4} {\bf v}_F\cdot{\bf q},
\end{equation}
and hence
\begin{equation}
{\rm Im}\chi_{yz}^{\rm inter}({\bf q},0)=2\sum_{\bf k}\delta f_{\bf k}\frac{\xi_{\bf k} \Delta}{E_{\bf k}^4} {\bf v}_F\cdot{\bf q}.
\end{equation}
Remarkably, the interband transitions perfectly cancel the intraband contribution regardless of the temperature, and we are left with
\begin{equation}
\label{eq:inter}
{\rm Im}\chi_{y z}^{QP}({\bf q},0)=0.
\end{equation}
Even though this result has been calculated to linear order in $q$ so as to highlight the delicate cancellation that nullifies the superconducting version of the adiabatic STT, it is feasible to obtain a concise analytical expression for ${\rm Im}\delta\chi_{y z}^{QP}({\bf q},0)$ to arbitrary order in $q$. 
The outcome reads
\begin{equation}
\label{eq:delta_ap}
{\rm Im}\,\delta \chi_{y z}^{QP}({\bf q},0)\simeq 2\sum_{\bf k} \delta f_{\bf k}\frac{E_{\bf k}\sin\theta_{\bf k}-E_{{\bf k-q}}\sin\theta_{\bf k-q}}{E_{\bf k}^2-E_{{\bf k-q}}^2} - ({\bf q}\to -{\bf q})= 2\sum_{\bf k} \delta f_{\bf k}\frac{\Delta_{\bf k}-\Delta_{{\bf k-q}}}{E_{\bf k}^2-E_{{\bf k-q}}^2} - ({\bf q}\to -{\bf q}),
\end{equation}
where we have allowed for a generic momentum-dependence in the superconducting order parameter.
For every value of ${\bf k}$ (i.e. for every quasiparticle), the numerator of Eq.~(\ref{eq:delta_ap}) contains the difference in the $x$-component of the effective ``magnetic'' field (i.e. $E_{\bf k}\sin\theta_{\bf k}$) before and after the quasiparticle scatters from ${\bf k}$ to ${\bf k-q}$.
A change in the $x$-component of the effective field during the quasiparticle scattering process indicates a change in the rate of precession of the order parameter.
When induced by a current, this change is the adiabatic STT.
 
In sum, the superconducting analogue of the adiabatic STT is nonzero only if the order parameter is momentum-dependent.
With the exception of Eq.~(\ref{eq:delta_ap}), we have limited ourselves to a momentum-independent order parameter throughout this paper.

\section {Superconducting analogue of the nonadiabatic STT}

In ferromagnets, the nonadiabatic STT term appearing in Eq.~(\ref{eq:lls}) emerges from the changes in ${\rm Im}\,\chi^{QP}_{\perp\perp}({\bf q},0)$ ($\perp=y,z$) that occur under transport currents, to first order in $q$. 
The starting expression for a clean superconductor is
\begin{equation}
{\rm Im}\delta\chi_{\perp\perp}^{QP}({\bf q},\omega)=\pi\sum_{{\bf k},\alpha,\beta} \delta f_{{\bf k}\alpha}\left[|\langle{\bf k}\alpha|\tau^\perp|{\bf k}-{\bf q}\beta\rangle|^2 \delta(E_{{\bf k}\alpha}-E_{{\bf k}-{\bf q}\beta}+\omega)-\left(\begin{array}{c} {\bf q}\to -{\bf q}\\\omega\to-\omega\end{array}\right)\right].
\end{equation}
In general, a proper theory of nonadiabatic STT would have to incorporate disorder vertex corrections along with transport perturbations.
This task, which remains to be completed in the magnetism community, is beyond the scope of the present work.
Here we include the finite quasiparticle lifetime only through a shift in the quasiparticle distributions, which is expected to be a reasonable approximation for $T_c\tau>1$.

For $\omega\ll\Delta$ (which is the regime of interest in the present work), only intraband ($\alpha=\beta$) transitions contribute.
Thus 
\begin{equation}
{\rm Im}\delta \chi_{\perp\perp}^{QP}({\bf q},0)=\pi\sum_{{\bf k}\alpha} \delta f_{{\bf k}\alpha}\left[|\langle{\bf k}\alpha|\tau^\perp|{\bf k}-{\bf q}\alpha\rangle|^2 \delta(E_{{\bf k}\alpha}-E_{{\bf k}-{\bf q}\alpha})-({\bf q}\to -{\bf q})\right].
\end{equation}
The Dirac delta can be manipulated as
\begin{equation}
\delta(E_{{\bf k}\alpha}-E_{{\bf k\mp q}\alpha})=\delta(\sqrt{\xi_{\bf k}^2+\Delta^2}-\sqrt{(\xi_{\bf k}\mp q v_F \cos\varphi)^2+\Delta^2})=\frac{E_{\bf k}}{|\xi_{\bf k}| q v_F}\left[\delta(\cos\varphi)+\delta\left(\cos\varphi\mp\frac{2\xi_{\bf k}}{q v_F}\right)\right],
\end{equation}
where $\varphi$ is the angle between ${\bf v}_F$ and ${\bf q}$.
The first term can be ignored because it eventually gives a vanishing contribution.
Accordingly,
\begin{equation}
{\rm Im}\delta\chi_{\perp\perp}^{QP}({\bf q},0)\simeq\pi\sum_{{\bf k}\alpha}\delta f_{{\bf k}\alpha}\frac{E_{\bf k}}{|\xi_{\bf k}| q v_F}\left[|\langle{\bf k}\alpha|\tau^\perp|{\bf k-q}\alpha\rangle|^2 \delta\left(\cos\varphi-\frac{2\xi_{\bf k}}{q v_F}\right)-|\langle{\bf k}\alpha|\tau^\perp|{\bf k+q}\alpha\rangle|^2\delta\left(\cos\varphi+\frac{2\xi_{\bf k}}{q v_F}\right)\right].
\end{equation}

Let us first discuss ${\rm Im}\delta\chi_{zz}^{QP}({\bf q},0)$.
We immediately see that it vanishes, because
\begin{equation}
\lim_{\cos\varphi\to \frac{2\xi_{\bf k}}{q v_F}} \langle {\bf k}\alpha|\tau^z|{\bf k- q}\alpha\rangle=\lim_{\cos\varphi\to -\frac{2\xi_{\bf k}}{q v_F}} \langle {\bf k}\alpha|\tau^z|{\bf k+ q}\alpha\rangle=0.
\end{equation}
Note that these relations follow from the exact eigenstates, without expanding in $q$.
An expansion in $q$ would be inappropriate in this case, because the delta function pins $\cos\varphi$ to $\sim 1/q$.

Next, we focus on ${\rm Im}\delta\chi_{yy}^{QP}({\bf q},0)$.
In this case,
\begin{equation}
\lim_{\cos\varphi\to\frac{2\xi_{\bf k}}{q v_F}} |\langle {\bf k}\alpha|\tau^y|{\bf k-q}\alpha\rangle|^2=\lim_{\cos\varphi\to-\frac{2\xi_{\bf k}}{q v_F}} |\langle {\bf k}\alpha|\tau^y|{\bf k+q}\alpha\rangle|^2=\frac{\xi_{\bf k}^2}{E_{\bf k}^2}
\end{equation}
and thus
\begin{align}
{\rm Im}\delta\chi_{yy}^{QP}({\bf q},0) &=\pi\sum_{{\bf k}\alpha}\delta f_{{\bf k}\alpha}\frac{\xi_{\bf k}^2}{E_{\bf k}^2}\frac{E_{\bf k}}{|\xi_{\bf k}| v_F q}\left[\delta\left(\cos\varphi-\frac{2\xi_{\bf k}}{q v_F}\right)-\delta\left(\cos\varphi+\frac{2\xi_{\bf k}}{q v_F}\right)\right]\nonumber\\
=& 2\pi \sum_{\bf k} \delta f_{\bf k} \frac{\xi_{\bf k}^2}{E_{\bf k}^2}\frac{E_{\bf k}}{|\xi_{\bf k}| v_F q}\left[\delta\left(\cos\varphi-\frac{2\xi_{\bf k}}{q v_F}\right)-\delta\left(\cos\varphi+\frac{2\xi_{\bf k}}{q v_F}\right)\right].
\end{align}
Since the expression multiplying $\delta f_{\bf k}$ is odd under $\xi_{\bf k}\to -\xi_{\bf k}$, a temperature gradient is required in order to obtain a nonzero result.
For such a perturbation, we plug in Eq.~(\ref{eq:rel}) and arrive at
\begin{equation}
\label{eq:res}
{\rm Im}\delta\chi_{yy}^{QP}({\bf q},0)\simeq 4\pi N_0 \tau v_F \frac{{\hat q}\cdot{\boldsymbol\nabla}T}{T}\frac{1}{(q v_F)^2}\int_{-\frac{q v_F}{2}}^{\frac{q v_F}{2}} d\xi \xi^2\frac{\partial f}{\partial E}\simeq -\pi N_0 D \frac{{\bf q}\cdot{\boldsymbol\nabla}T}{T}\frac{1}{4 T \cosh^2\frac{\Delta}{2 T}}.
\end{equation}
Although this equation has been derived for three dimensions, we have verified by explicit calculation that the final result is valid for two dimensions as well.
In the 2D case, one must use
\begin{equation}
\int_{-q v_F/2}^{q v_F/2}d\xi \xi^2\frac{q v_F}{\sqrt{q^2 v_F^2-4\xi^2}}\simeq \frac{\pi}{16} (q v_F)^3. 
\end{equation}

In this Appendix, as in the previous one,  we have used $\xi_{\bf k-q}=\xi_{\bf k}-{\bf k}\cdot{\bf q}/m+q^2/(2 m)\simeq \xi_{\bf k}-{\bf v}_F\cdot{\bf q}$.
It can be shown that keeping the $q^2$ term in this expansion (which amounts to breaking particle-hole symmetry) will result in small ($\propto q/k_F$) nonzero values for ${\rm Im}\delta\chi_{\perp\perp}$ in presence of a transport perturbation that is {\em even} under $\xi_{\bf k}\to -\xi_{\bf k}$.
For example, we find that a uniform electric field ${\bf E}$ leads to 
\begin{equation}
{\rm Im}\delta\chi_{yy}^{QP}({\bf q},0)\propto N_0\frac{q v_F}{T}\frac{e l {\hat q}\cdot{\bf E}}{\mu},
\end{equation}
which is $O(T/\mu)$ smaller than Eq.~(\ref{eq:res}) for a fixed strength of the perturbation.

\section{Transport-induced changes in the dynamical anisotropy field}

In the simplest toy models for itinerant magnets, where intrinsic spin-orbit coupling is ignored, the real part of $\chi^{QP}_{\perp\perp}$ does not change under a transport perturbation.
%; instead, only the imaginary part of $\chi^{QP}_{z y}$ is altered.
However, even the simplest toy model for superconductivity has some intrinsic ``pseudospin-orbit coupling'', because the kinetic energy of electrons acts as a momentum-dependent magnetic field in particle-hole (Nambu) space.
With this in mind, we evaluate $\chi_{yy}^{QP}$ and $\chi_{zz}^{QP}$ with shifted quasiparticle distribution functions.
The starting point is
\begin{equation}
{\rm Re}\,\delta\chi_{\perp\perp}^{QP}({\bf q},\omega)=-\sum_{{\bf k}\alpha\beta}\delta f_{{\bf k}\alpha}\left[\frac{|\langle{\bf k}\alpha|\tau^\perp|{\bf k}-{\bf q}\beta\rangle|^2}{E_{{\bf k}\alpha}-E_{{\bf k-q}\beta}+\omega}+\frac{|\langle{\bf k}\alpha|\tau^\perp|{\bf k}+{\bf q}\beta\rangle|^2}{E_{{\bf k}\alpha}-E_{{\bf k+q}\beta}+\omega}\right].
\end{equation}
Let us begin from the intraband contributions for $\perp\perp=zz$:
\begin{equation}
{\rm Re}\,\delta\chi_{zz}^{\rm intra}({\bf q},\omega)\simeq -\sum_{{\bf k}\alpha}\delta f_{{\bf k}\alpha}\frac{[|\langle{\bf k}\alpha|\tau^z|{\bf k}-{\bf q}\alpha\rangle|^2-({\bf q}\to -{\bf q})]}{{\bf v}_{{\bf k}\alpha}\cdot{\bf q}+\omega}
- \frac{1}{2}\sum_{{\bf k}\alpha} \delta f_{{\bf k}\alpha}\frac{({\bf q}\cdot\vec\nabla_{\bf k})^2 E_{{\bf k}\alpha}}{(\omega+{\bf v}_{{\bf k}\alpha}\cdot{\bf q})^2}\left[\langle {\bf k}\alpha|\tau^z|{\bf k-q}\alpha\rangle|^2 + ({\bf q}\to -{\bf q})\right].
\end{equation}
Expanding the terms inside the square brackets to lowest order in momentum and recalling that $\delta f_{{\bf k}+}=\delta f_{{\bf k}-}$, we arrive at
\begin{equation}
{\rm Re}\,\delta\chi_{zz}^{\rm intra}({\bf q},\omega)\simeq 4 \sum_{\bf k} \delta f_{{\bf k}} \frac{\omega {\bf v}_F\cdot{\bf q}}{\omega^2-({\bf v}_{\bf k}\cdot{\bf q})^2}\frac{\omega^2}{\omega^2-({\bf v}_{\bf k}\cdot{\bf q})^2}\frac{\Delta^2 \xi_{\bf k}}{E_{\bf k}^4}.
\end{equation}
Next, let us look at the interband contribution.
In the low-frequency and long wavelength expansion, 
\begin{align}
 {\rm Re}\delta\chi_{zz}^{\rm inter}({\bf q},\omega)&\simeq-\sum_{{\bf k}\alpha\neq\beta} \delta f_{{\bf k}\alpha}\frac{|\langle {\bf k}\alpha|\tau^z|{\bf k}-{\bf q}\beta\rangle|^2}{E_{{\bf k}\alpha}-E_{{\bf k}\beta}}\left[1-\frac{\omega+{\bf v}_{{\bf k}\beta}\cdot{\bf q}}{E_{{\bf k}\alpha}-E_{{\bf k}\beta}}\right] +\left(\begin{array}{c} {\bf q}\to -{\bf q} \\ \omega\to-\omega\end{array}\right)\nonumber\\
&\simeq -4\sum_{{\bf k}\alpha\neq\beta}\delta f_{{\bf k}\alpha}\frac{\omega+{\bf v}_{{\bf k}\beta}\cdot{\bf q}}{(E_{{\bf k}\alpha}-E_{{\bf k}\beta})^2}\langle{\bf k}\alpha|\tau^z|{\bf k}\beta\rangle\langle {\bf k}\alpha|\tau^z {\bf q}\cdot\nabla_{\bf k}|{\bf k}\beta\rangle.
\end{align}
where we have used $\delta f_{\bf k}=-\delta f_{-\bf k}$.
Computing the matrix elements, we get
\begin{equation}
{\rm Re}\,\delta\chi_{zz}^{\rm inter}({\bf q},\omega)\simeq \sum_{\bf k} \delta f_{\bf k}\frac{\omega ({\bf q}\cdot {\bf v}_F)}{E_{\bf k}^2}\frac{\xi_{\bf k}\Delta^2}{E_{\bf k}^4}.
\end{equation}
%Note that, in the long-wavelength and low-frequency regime, the intraband part is parametrically smaller than the interband part.
%Therefore, we can approximate
%\begin{equation}
%{\rm Re}\,\delta\chi_{zz}^{QP}({\bf q},\omega)\simeq -4\sum_{\bf k}\delta f_{\bf k}\frac{\omega ({\bf q}\cdot {\bf v}_F)}{({\bf v}_{\bf k}\cdot{\bf q})^2-\omega^2}\frac{\xi_{\bf k%}\Delta^2}{E_{\bf k}^4}
%\end{equation}

One may compute ${\rm Re}\delta\chi_{yy}^{QP}$ following identical steps.
The final result is
\begin{equation}
{\rm Re}\,\delta\chi_{yy}^{QP}({\bf q},\omega)\simeq 2\sum_{\bf k}\delta f_{\bf k}\frac{\omega ({\bf q}\cdot {\bf v}_F)}{({\bf v}_{\bf k}\cdot{\bf q})^2-\omega^2}\frac{\xi_{\bf k}\Delta^2}{E_{\bf k}^4}\frac{({\bf q}\cdot{\bf v}_F)^2}{E_{\bf k}^2},
\end{equation}
where we have neglected the interband contribution (which is parametrically smaller) and have also ignored terms that are $\sim O(\Delta^2/T^2)$ smaller.
Although ${\rm Re}\delta\chi_{yy}^{QP}$ is of higher order in $q$ than ${\rm Re}\delta\chi_{zz}^{QP}$,
it can make a contribution of the same order to the collective mode frequency (the reason being that the $yy$ sector of the response function is $\sim O(q^2,\omega^2)$, while the $zz$ sector contains a term that does not vanish at $\omega\to 0$ and $q\to 0$).

Once again we observe that only transport perturbations that are odd under $\xi_{\bf k}\to -\xi_{\bf k}$ (e.g. a temperature gradient) will lead to a nonzero ${\rm Re}\,\delta\chi_{\perp\perp}^{QP}$.
In addition, the above expressions indicate that the transport correction to the dynamical anisotropy field contains two distinct regimes: $\omega\gg v_F q$ and $\omega\ll q v_F$.
In the regime $\omega\gg q v_F$, we obtain
\begin{equation}
\label{eq:zz_an}
{\rm Re}\,\delta\chi_{zz}^{QP} ({\bf q},\omega)\simeq 4 \sum_{\bf k}\delta f_{\bf k}\frac{{\bf q}\cdot{\bf v}_F}{\omega}\frac{\Delta^2 \xi_{\bf k}}{E_{\bf k}^4}\simeq-\frac{8}{\pi} N_0 I D \frac{{\bf q}\cdot{\boldsymbol\nabla} T}{\omega T}, 
\end{equation} 
whereas the contribution from ${\rm Re}\delta\chi_{yy}^{QP}$ to the collective mode dispersion can be safely neglected.
In the opposite regime, $\omega\ll q v_F$, we have
\begin{equation}
{\rm Re}\delta\chi_{yy}^{QP}({\bf q},\omega)\simeq \frac{1}{6} N_0 D\frac{{\bf q}\cdot{\boldsymbol\nabla}T}{T^2}\frac{\omega}{\Delta},
\end{equation}
whereas the contribution from ${\rm Re}\delta\chi_{zz}^{QP}$ to the collective mode dispersion can be safely neglected.

\section{Amplitude fluctuations remain decoupled when $\nabla T\neq 0$}

In the main text we have discussed how the phase-charge fluctuations are altered by transport perturbations.
This change is the superconducting analogue of the spin transfer torque.
Ignoring small departures from particle-hole symmetry, we have found that perturbations leading to an electrical current do not change the phase-charge coupling, while perturbations leading to a heat current do change it.
One may have the concern that applying a temperature gradient could result in the coupling between amplitude and phase/charge fluctuations.
Here we show that not to be the case.

We begin determining $\chi_{x y}^{QP}$ in presence of drifted quasiparticle factors:
\begin{equation}
\chi_{y x}^{QP}({\bf q},\omega)=-\sum_{{\bf k}\alpha\beta} (f_{{\bf k}\alpha}-f_{{\bf k-q}\beta})\frac{\langle{\bf k}\alpha|\tau^y|{\bf k-q}\beta\rangle\langle{\bf k-q}\beta|\tau^x|{\bf k}\alpha\rangle}{E_{{\bf k}\alpha}-E_{{\bf k-q}\beta}+\omega+i 0^+}.
\end{equation}
Recognizing that $\langle{\bf k}\alpha|\tau^y|{\bf k-q}\beta\rangle\langle{\bf k-q}\beta|\tau^x|{\bf k}\alpha\rangle$ is purely imaginary,
\begin{equation}
\chi_{y x}^{QP}({\bf q},\omega)=-\sum_{{\bf k}\alpha\beta} f_{{\bf k}\alpha}\left[\frac{\langle{\bf k}\alpha|\tau^y|{\bf k-q}\beta\rangle\langle{\bf k-q}\beta|\tau^x|{\bf k}\alpha\rangle}{E_{{\bf k}\alpha}-E_{{\bf k-q}\beta}+\omega+i 0^+} -\left(\begin{array}{c} 0^+\to -0^+\\ \omega\to -\omega\\ {\bf q}\to -{\bf q}\end{array}\right)\right].
\end{equation}
First, the real part reads
\begin{equation}
{\rm Re}\,\chi_{y x}^{QP}({\bf q},\omega)= i\pi\sum_{{\bf k}\alpha\beta} f_{{\bf k}\alpha}\left[\langle{\bf k}\alpha|\tau^y|{\bf k-q}\beta\rangle\langle{\bf k-q}\beta|\tau^x|{\bf k}\alpha\rangle \delta(E_{{\bf k}\alpha}-E_{{\bf k-q}\beta})+({\bf q}\to -{\bf q})\right].
\end{equation}
Because the term inside the square is even under ${\bf k}\to -{\bf k}$, ${\rm Re}\chi_{y x}^{QP}({\bf q},\omega)$ remains unchanged (i.e. zero) under a transport perturbation.
Next, consider the imaginary part. 
The leading order contribution comes from
\begin{equation}
{\rm Im}\chi_{y x}^{QP}({\bf q},0)=-\sum_{{\bf k}\alpha\beta} f_{{\bf k}\alpha}\left[\frac{{\rm Im}[\langle{\bf k}\alpha|\tau^y|{\bf k-q}\beta\rangle\langle{\bf k-q}\beta|\tau^x|{\bf k}\alpha\rangle]}{E_{{\bf k}\alpha}-E_{{\bf k-q}\beta}}-({\bf q}\to -{\bf q})\right],
\end{equation}
which may be calculated exactly in the same way as ${\rm Im}\chi_{y z}^{QP}({\bf q},0)$; the only difference comes from the overlap matrix elements.
The final result is
\begin{equation}
{\rm Im}\chi_{y x}^{QP}({\bf q},0)\simeq\sum_{\bf k}\delta f_{\bf k}\frac{1}{\xi_{\bf k}^2} {\bf q}\cdot{\bf v}_F
\end{equation}
Because the factor multiplying $\delta f_{\bf k}$ is even under $\xi_{\bf k}\to -\xi_{\bf k}$, a temperature gradient will {\em not} lead to any change in ${\rm Im}\chi_{y x}^{\rm intra}({\bf q},0)$ (i.e., it will remain zero).
In contrast, a transport perturbation that is even under $\xi_{\bf k}\to -\xi_{\bf k}$ (i.e. a perturbation that creates an electrical current) would lead to ${\rm Im}\chi^{\rm intra}_{y x}({\bf q},\omega)\neq 0$.

A straightforward evaluation of $\chi_{z x}^{QP}$ leads to an identical conclusion, namely that a temperature gradient does not induce a coupling between amplitude and charge/phase fluctuations irrespective of the temperature of the system.
It is interesting that an electric current couples amplitude fluctuations with charge/phase fluctuations but does not directly alter the coupling between charge and phase fluctuations (i.e. it induces no STT), whereas a heat current does exactly the opposite.

\section{Collective modes in ultraclean superconductors}

In the main text we have shown the influence of the superconducting STT in the response functions of disordered superconductors with $\tau^{-1}\gg\omega$.
For completeness, here we discuss clean superconductors, where $\omega\gg\tau^{-1}$, even though in practice this condition is difficult to satisfy at subgap frequencies. 
For uniform temperature, the charge/phase response of a 3D superconductor near $T_c$ reads~\cite{wong1988}
\begin{equation}
\left(\begin{array}{cc} \frac{\omega^2}{2\Delta^2} I -\frac{1}{2\Delta^2}\frac{n_s}{n}\frac{v_F^2 q^2}{3}+i {\rm Im}\chi_{yy}^{QP}/N_0 & i\frac{\omega}{\Delta}\frac{n_s}{n}\\
                              -i\frac{\omega}{\Delta}\frac{n_s}{n} &  2+\frac{1}{N_0 V_{\bf q}} + i {\rm Im}\chi_{zz}^{QP}/N_0
\end{array}\right)\left(\begin{array}{c} \delta\Delta^y \\ e\phi\end{array}\right)=0,
\end{equation}
where $I=\pi\Delta/(4 T)$ and $V_{\bf q}=4\pi e^2/q^2$.
In the derivation of this result we have used
\begin{equation}
\int d\xi\left(\frac{-\partial f}{\partial E}\right) \int\frac{d\Omega_{\bf k}}{4\pi}\frac{({\bf v}_{\bf k}\cdot{\bf q})^2}{({\bf v}_{\bf k}\cdot{\bf q})^2-\omega^2}
\simeq  \int d\xi\left(\frac{-\partial f}{\partial E}\right) \int\frac{d\Omega_{\bf k}}{4\pi}\frac{({\bf v}_F\cdot{\bf q})^2}{({\bf v}_F\cdot{\bf q})^2-\omega^2}\simeq\int d\xi\frac{-\partial f}{\partial E},
\end{equation}
where we have recognized that $-\partial f/\partial E=1/(4 T \cosh^2 (E/2T))$, which for $T\simeq T_c$ (i.e. $T\gg \Delta$) limits the main contribution of the integrand to $\xi\simeq T$ (note that the $\xi\ll E$ regime is depleted by the factor $v_k^2$ in the numerator).
Consequently, $v_k=(\xi_k/E_k) v_F\simeq v_F$.
Moreover, we have anticipated that $\omega\ll v_F q$.

Without the damping terms, the collective mode dispersion reads 
\begin{equation}
\label{eq:av}
\omega_\pm ({\bf q})=\pm\sqrt{\frac{n_s}{3 I n}} v_F q=\pm\left(\frac{7\zeta(3)}{3\pi^3}\frac{\Delta}{T}\right)^{1/2} v_F q.
\end{equation}
Note that this mode is essentially a phase-only mode, in which the phase-charge coupling has been neglected.
Since $\omega < v_F q$, one needs to consider the Landau damping.
On one hand,
\begin{equation}
\label{eq:landau}
{\rm Im}\chi_{yy}^{QP}({\bf q},\omega)=\pi\sum_{\bf k}(f_{{\bf k}\alpha}-f_{{\bf k-q}\alpha})|\langle{\bf k}\alpha|\tau^y|{\bf k-q}\alpha\rangle|^2\delta(E_{{\bf k}\alpha}-E_{{\bf k-q}\alpha}+\omega)\simeq\pi N_0\frac{v_F q}{4 T}\frac{\omega}{4\Delta},
\end{equation}
where we have used $q v_F\ll\Delta\ll T$.
Due to Eq.~(\ref{eq:landau}), the above collective mode becomes overdamped and thus hardly observable.
Incidentally, the Landau damping term of the charge sector, ${\rm Im}\chi_{zz}^{QP}({\bf q},\omega)\propto \omega/(q v_F)$, plays no role in the dispersion of the collective mode.

In presence of a temperature gradient, the influence of the nonadiabatic STT term is to modify the Landau damping.
A priori, there is the intriguing possibility that the STT term may cancel the Landau damping (first along the direction of momentum {\bf q} that is parallel or antiparallel to ${\boldsymbol\nabla}T$) and thus render a propagating collective mode.
However, for experimentally reasonable temperature gradients, the STT term is parametrically smaller than the Landau damping term (due to $v_F\gg v_T$) and thus the collective mode will remain overdamped. 

For 2D superconductors, the response function obeys
\begin{equation}
\left(\begin{array}{cc}\frac{\omega^2}{2\Delta^2}I -\frac{1}{2\Delta^2}\frac{n_s}{n}\frac{v_F^2 q^2}{2} & i\frac{\omega}{\Delta}I\\
                                   -i\frac{\omega}{\Delta}I & 2\left(I -\frac{v_F^2 q^2}{2\omega^2}\right)-\frac{q}{2\pi e^2 N_0}
\end{array}\right)\left(\begin{array}{c}\delta\Delta^y\\ e\phi\end{array}\right)=0.
\end{equation}
In the derivation of this equation we have used
\begin{equation}
\int d\xi\frac{\Delta^2}{E^2}\left(\frac{-\partial f}{\partial E}\right)\int_0^{2\pi}\frac{d\varphi}{2\pi} \frac{({\bf v}_F\cdot{\bf q})^2}{({\bf v}_F\cdot{\bf q})^2-\omega^2}\nonumber\\
\simeq-\frac{v_F^2 q^2}{2\omega^2}\int d\xi\frac{\Delta^2}{E^2}\left(\frac{-\partial f}{\partial E}\right)\simeq -\frac{v_F^2 q^2}{2\omega^2}\left(I-\frac{n_s}{n}\right),
\end{equation}
where in the first equality we have anticipated that $\omega = c q^{1/2}\gg v_F q$ at $q\ll 2\pi e^2 N_0$ [for $q\gg 2\pi e^2 N_0$ one simply recovers the 2D version of Eqs.~(\ref{eq:av}) and (\ref{eq:landau})], and in the second equality we have referred to Ref.~[\onlinecite{wong1988}].
In this regime, the Landau damping is absent.
Consequently, the collective mode dispersion is
\begin{align}
\omega_\pm({\bf q}) & =\pm\sqrt{2\pi e^2 N_0 v_F^2 q} =\pm\sqrt{4\pi e^2 n q /m} , %,\,\,\, \mbox{(2D plasmon)} \mbox{    for    } q\ll 2\pi e^2 N_0\nonumber\\
\end{align}
i.e. the ordinary 2D plasmon of metals (note the difference with respect to the disordered case discussed in the main text, where the plasmon frequency contained $n_s$ instead of $n$) .

A temperature gradient modifies the 2D plasmon.
However, Eq.~(\ref{eq:yz_1}) is not accurate for the evaluation of the collective mode dispersion in the $\omega\gg v_F q$ regime.
In this frequency regime, the nonadiabatic STT vanishes (for the same phase space reason for which the Landau damping vanishes).
However, there is a non-vanishing transport contribution that originates from the interband part of $\delta\chi_{yz}^{QP}$ (the intraband part is depleted in this regime) as well as from the dynamical anisotropy field (cf. Appendix C).
Using Eqs.~(\ref{eq:inter}) and (\ref{eq:zz_an}), we arrive at
\begin{equation}
\label{eq:pl_clean}
\omega_\pm({\bf q})=-\delta\omega\pm\sqrt{4\pi e^2 n q /m +\delta\omega^2},\,\,\,\,\mbox{where}\,\,\,\,\delta\omega=8 e^2 N_0 I D \frac{{\hat q}\cdot{\boldsymbol\nabla}T}{T}.
\end{equation}
Hence, for a 2D plasmon with $\omega\gg q v_F$, the real part of the dispersion is changed by driving BCS quasiparticles out of equilibrium.
It must be noted that the contributions from Eqs.~(\ref{eq:inter}) and (\ref{eq:zz_an}) partly cancel each other; however, we have not found a perfect cancellation.
%In Eq.~(\ref{eq:pl_clean}) we have disregarded the contribution from $\delta\chi^{QP}_{yz}({\bf q},\omega)$, which is no longer zero for $\omega\gg q v_F$.
%This omission does not qualitatively change the result above.
\end{widetext}


\begin{thebibliography}{100}
\bibitem{units} In Gaussian units. In order to translate to SI units, we replace $e^2\to e_0^2/(4\pi\epsilon)$, where $e_0$ is the electron charge and $\epsilon$ is the dielectric constant.
\bibitem{anis} In a ferromagnet with magnetization ${\bf M}=M\hat{\Omega}$, the anisotropy field is given by ${\bf H}_{{\rm an}}^i=-M [\chi^{-1}({\bf 0},0)]_{i j} \hat{\Omega}_j$, where $\chi^{-1}=(\chi^{QP})^{-1}-U$. 
Similarly, the exchange field is given by ${\bf H}_{\rm xc}^i=-M [\chi^{-1}({\bf q},0)-\chi^{-1}({\bf 0},0)]_{i j} \hat{\Omega}_j$.
\bibitem{caveat0} Our approximation is partly informed by the transport equation description of nonequilibrium superconductivity.\cite{aronov} In Ref.~[\onlinecite{aronov}], the nonequilibrium contribution to one-body observables (e.g. charge and current densities) are evaluated by writing the one-body operator for a clean system, and incorporating finite lifetimes solely through shifted quasiparticle distribution functions. 
Our approximation mimics this strategy for two-body observables (response functions).
\end{thebibliography}
\end{document}